\documentclass{article}
\usepackage{caption}
\usepackage{prime}
\usepackage[utf8]{inputenc} 
\usepackage{amsfonts}       
\usepackage{nicefrac}       
\usepackage{graphicx}
\usepackage{float}
\usepackage{url}
\usepackage[export]{adjustbox}
\pdfoutput=1
\usepackage{hyperref}
\usepackage{titling} 
\usepackage{multicol} 

\title{{A versatile machine learning workflow for 
		high-throughput analysis of supported metal catalyst particles
}}

\author{
	Arda Genc$^{1}$, Justin Marlowe$^{2}$, Anika Jalil$^{2}$, Libor Kovarik$^{3}$, Phillip Christopher$^{2}$
}

\date{} 

\begin{document}
	\maketitle
	\vspace{-4em} 
	\begin{center}
		$^{1}$Materials Department, University of California Santa Barbara, Santa Barbara, CA\\
		$^{2}$Department of Chemical Engineering, University of California Santa Barbara, Santa Barbara, CA\\
		$^{3}$Institute for Integrated Catalysis, Pacific Northwest National Laboratory, Richland, WA\\
	\end{center}
	\vspace{3em} 
\begin{abstract}
Accurate and efficient characterization of nanoparticles (NPs), particularly regarding particle size distribution, is essential for advancing our understanding of their structure-property relationship and facilitating their design for various applications. In this study, we introduce a novel two-stage artificial intelligence (AI)-driven workflow for NP analysis that leverages prompt engineering techniques from state-of-the-art single-stage object detection and large-scale vision transformer (ViT) architectures. This methodology was applied to transmission electron microscopy (TEM) and scanning TEM (STEM) images of heterogeneous catalysts, enabling high-resolution, high-throughput analysis of particle size distributions for supported metal catalysts. The model’s performance in detecting and segmenting NPs was validated across diverse heterogeneous catalyst systems, including various metals (Cu, Ru, Pt, and PtCo), supports (silica ($SiO_2$), $\gamma$-alumina ($\gamma$-$Al_2O_3$), and carbon black), and particle diameter size distributions with mean and standard deviations of 2.9 $\pm$ 1.1 nm, 1.6 $\pm$ 0.2 nm, 9.7 $\pm$ 4.6 nm, 4 $\pm$ 1.0 nm.  Additionally, the proposed machine learning (ML) approach successfully detects and segments overlapping NPs anchored on non-uniform catalytic support materials, providing critical insights into their spatial arrangements and interactions. Our AI-assisted NP analysis workflow demonstrates robust generalization across diverse datasets and can be readily applied to similar NP segmentation tasks without requiring costly model retraining.
\end{abstract}
\section{Introduction}
Heterogeneous catalysts, consisting of metal/oxide nanoparticles (NPs) deposited on high-surface-area supports, are fundamental for major industrial processes and environmental management. Their widespread use is driven by their small size and high surface area-to-volume ratios, which endow them with distinctive physical and chemical properties and efficient metal utilization. In heterogeneous catalysis, precious metal NPs such as Platinum (Pt) and Palladium (Pd) are extensively employed as active sites for various chemical reactions owing to their unique surface electronic structure that facilitates a range of important catalytic processes  \cite{WHEELDON2017127,Friend2017}. 

Accordingly, accurate analysis of NP size distributions is critical for understanding the structure-property relationship of heterogeneous catalysts. Several techniques are commonly employed to measure NP size, including  $H_2$ or CO chemisorption, dynamic light scattering (DLS) for NPs in solution, TEM, scanning electron microscopy (SEM), X-ray diffraction (XRD), nanoparticle tracking analysis (NTA), small-angle X-ray scattering (SAXS), X-ray absorption spectroscopy, UV-visible spectroscopy, etc. \cite{Altammar2023, Caputo2019,Matyi2020}. Each method has distinct advantages and disadvantages. A comprehensive review of these techniques is beyond the scope of our paper.

Among these techniques, TEM employs an electron beam to probe thin samples, yielding high-resolution images that reveal the detailed structure and morphology of materials \cite{Smith2015}.  One of TEM's key strengths is its ability to achieve sub-nanometer resolution, allowing for direct observation of particle shape and structure. This high-level detail is essential for many analysis tasks, which require precise identification of objects and their boundaries at pixel resolutions of tenths of a nanometer and even at subpixel precision.

While providing high-resolution information, TEM is not a sample-averaged technique; it focuses on a limited area of a sample by acquiring 2D projections of 3D objects. Due to the lack of automation in data acquisition and analysis, TEM is often overlooked for large-scale, routine analysis, which hinders its broader application {\cite{Pratiush2024}. As the field progresses toward scaling 2D data analysis, there is a growing demand for reliable, reproducible, and high-throughput methods for particle size analysis in TEM \cite{Xiang1995,Christopher2020,Meyer2024}. Efficiently processing large datasets is crucial for obtaining statistically meaningful information, particularly when a single TEM grid can contain tens of thousands of NPs. Addressing these challenges is essential for advancing the quantitative analysis of NP systems and enhancing the efficiency of TEM-based studies.

Classical image processing methods such as thresholding, edge detection, and watershed segmentation, commonly employed in NP size analysis, use global intensity distributions rather than localized pixel-level classifications for feature extraction \cite{Plaksyvyi2023,Khaniabadi2024,Jena2018,kousaka2024}. These methods often fail when analyzing complex, densely packed, low-contrast, and fine-scale NPs when scattered on catalytic support materials with similar contrast. In such cases, manual user intervention is required to resolve objects and their boundaries. Without this, image analysis results in ambiguities due to the intricate contrast mechanisms inherent to TEM.

Recent advances in AI have profoundly transformed the field of computer vision, enabling the automated analysis of large and complex datasets that would otherwise be impractical manually
\cite{lecunDeepLearning2015,krizhevskyImageNetClassificationDeep2012,longFullyConvolutionalNetworks2015,ciresanDeepNeuralNetworks2012,Genc2022,decostHighThroughputQuantitative2019,greenwaldWholecellSegmentationTissue2021,robertsDeepLearningSemantic2019,heDelvingDeepRectifiers2015, horwathUnderstandingImportantFeatures2020}. Deep learning-based pixel-level classifiers have been effectively employed in various classification tasks where traditional methods struggle to process complex structures  \cite{akersRapidFlexibleSegmentation2021,Ziatdinow2017,Lakshmipriya2023}. AI-driven approaches excel in cases where bimodal intensity distributions are ineffective, and boundaries between the structures are difficult to resolve due to the subtle variations in the contrast gradients.

Convolutional neural networks (CNNs) have been explored for the segmentation of NPs in high-resolution TEM images (HRTEM)\cite{horwathUnderstandingImportantFeatures2020, D2NA00781A, Groschner_Choi_Scott_2021}. These studies primarily focus on segmenting HRTEM images of well-isolated precious metal NPs dispersed on ultrathin amorphous carbon film TEM supports. While the CNNs have been evaluated on these ideal systems, their applicability has not been assessed for supported metal catalyst particles, where the background intensity of the support can exhibit significant complexity. Furthermore, these approaches have shown limited success in segmenting overlapped or connected particles, as well as fine-scale particles with diameters of 2 nm or less.

In AI-driven analysis, supervised and unsupervised machine learning (ML) algorithms face challenges regarding model generalization and compatibility \cite{Yuan2023,Holm2020}. CNNs, such as U-Net, typically require large amounts of labeled data for model training to achieve optimal performance  \cite{horwathUnderstandingImportantFeatures2020, D2NA00781A, ronnebergerUNetConvolutionalNetworks2015}. Their effectiveness diminishes when applied to images that differ from those in the training dataset, limiting their ability to generalize. Consequently, ensuring these models perform consistently across diverse datasets remains a critical challenge in advancing AI-driven analysis. 

To overcome the challenges associated with NP detection and analysis, we developed a two-stage methodology that integrates an object detection stage, functioning as a prompt generator and a transformer model for pixel-level classification. Our NP analysis approach encompasses Meta AI's segment anything model (SAM), which is designed to function through manual or automated prompts \cite{Kirillow2023}. As a foundation ViT model, SAM generalizes effectively across various segmentation tasks. Implementing models like SAM represents a paradigm shift in deep-learning-based image segmentation, as traditional deep-learning approaches typically involve training CNNs from scratch or fine-tuning them for specific tasks. Recent studies have reported significant segmentation accuracy gains over traditional CNN models like U-Net, leveraging SAM's zero-shot generalization capability by point or box prompt engineering strategies \cite{Colber2024, Wu2023, Wang2023, nichols2024segmentmodelgraincharacterization,rafaeli2024promptbasedsegmentationmultipleresolutions}.

Object detection consists of two competing domains: single-stage and two-stage detectors for bounding box predictions. Two-stage models, such as Faster R-CNN, first generate a set of region proposals for potential object locations \cite{Ren2016}. These proposals are then refined and classified into object categories in a second stage. In contrast, single-stage object detection models process the entire image simultaneously, directly predicting bounding boxes without a separate proposal stage, making them faster and more suitable for real-time applications  \cite{Redmon2016}. Among two-stage detectors, Faster R-CNN is a leading model, excelling in accuracy. However, single-stage models like YOLO (You Only Look Once), particularly with the development of YOLOv8, have significantly improved their architectural design and model generalization through anchor-free and multi-scale prediction algorithms \cite{Jocher2023, tian2019fcosfullyconvolutionalonestage}. These models have been reported to be comparable to, and even outperform, Faster R-CNN in certain object detection tasks  \cite{Ezzeddini2024,Bery2024,Sapkota2024,duan2019centernetkeypointtripletsobject}.

Single-stage object detection models like YOLO perform well in rapidly and accurately localizing objects across various sizes and with occlusions \cite{Liu2018,Terven2023}. YOLO's robustness in handling these complex visual tasks has established its versatility in numerous applications, including autonomous vehicle navigation, aerial drone surveillance, defect identification in industrial manufacturing, and even early-stage cancer detection  \cite{Haifawi2023,Jia2023,Baccouche2022,Li2018}. 

Our approach combines YOLO's compatibility, speed, and accuracy for precise NP detection with SAM’s fast and accurate segmentation capabilities, providing a robust tool for high-throughput analysis. This integrated process demonstrates strong generalizability, eliminating the need for labor-intensive model retraining and allowing users to focus on data interpretation and analysis. We believe this approach will significantly accelerate TEM particle analysis, enabling the efficient processing of thousands of NPs with minimal user intervention and facilitating the automated analysis of large datasets.

\section{Results and discussion}
\subsection{Detection and segmentation}
Model training for the object detection task begins with transfer learning using the pre-trained weights of the YOLOv8x model, which serves as the foundation for the prompt generator. This model is then fine-tuned using manually annotated images to adapt it specifically for the NP class. This fine-tuning step enables the model to effectively detect NPs within bright-filed TEM (BF-TEM) and high-angle annular dark-field (HAADF) STEM images.

We selected the largest version of the YOLOv8 model, known as YOLOv8x (extra-large), which features more layers (365 layers), channels, and parameters (68M parameters). This architecture enhances the model's ability to detect smaller objects and manage complex scenes, making it particularly well-suited for NP detection.

Figure 1 (top) illustrates the three main components of the YOLO object detection model. The backbone, comprising a 53-layer deep convolutional neural network, is responsible for extracting feature maps from input images. The neck of the network integrates features from various stages of the backbone to produce a multi-scale feature representation. employing a path aggregation network (PANet) to enhance feature aggregation. The head of the network is responsible for predicting bounding boxes, objectness scores, and class probabilities. By utilizing feature maps from different levels of the network, the model can predict objects of varying sizes, ranging from small to large.

\begin{figure}[h]
	\centering
	\includegraphics[width=0.8\textwidth]{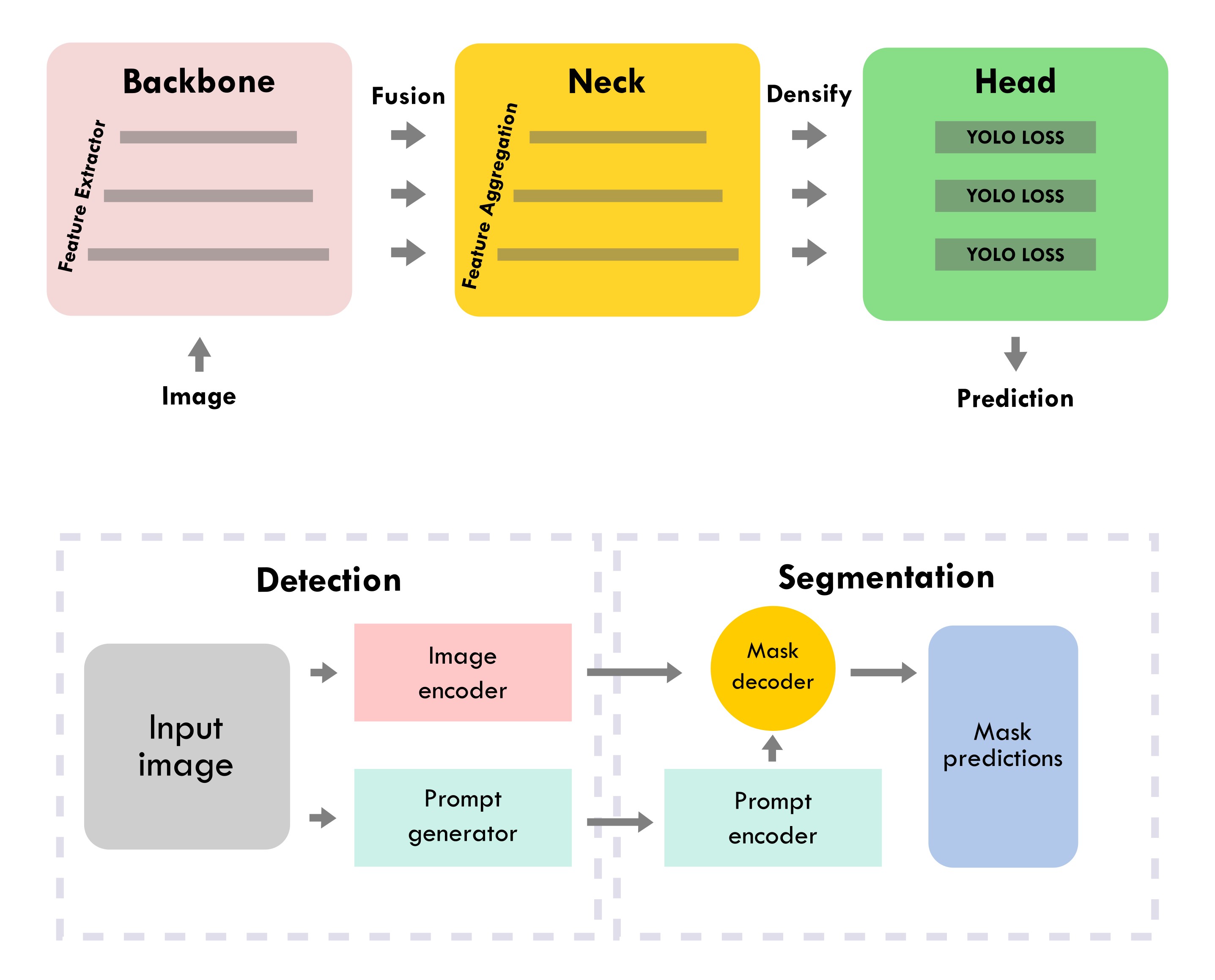}
	\caption{\small Schematic illustrating the main components of  YOLO architecture (top) and the proposed two-stage AI-assisted TEM particle analysis workflow, consisting of detection and segmentation stages (bottom).}
	\label{fig:fig1}
\end{figure}
In conjunction with YOLO, our pipeline encompasses SAM, which has been trained on the SA-1B dataset, consisting of 11 million images and 1 billion annotations. SAM has shown strong zero-shot learning capabilities, enabling segmentation without requiring additional retraining for new classes. We adapted the model using the smallest backbone, $\textit{vit-b}$, which provides a lightweight solution for fast and accurate pixel-level classification of the NPs. 

As illustrated in Figure 1 (bottom), we propose a two-stage workflow for AI-assisted particle analysis, consisting of detection and segmentation stages. The NP analysis process begins by acquiring BF-TEM and HAADF STEM images, which reveal NPs with varying shapes, sizes, and contrast levels on catalytic support materials. These images are processed by the SAM's image encoder, which generates feature representations and image embeddings that capture rich contextual information. The prompt generator localizes individual NPs within the TEM images, generating a bounding box for each NP detection. The bounding box predictions from the prompt generator are passed as inputs to the SAM's prompt encoder. The mask decoder then predicts multiple segmentation masks for each prompt by utilizing both the bounding box prompts and the image embeddings. The segmentation masks with the highest confidence scores are selected for NP analysis. The final segmentation stage allows for the extraction of key morphological and dimensional characteristics, such as NP size, shape, and area, on the catalytic support material. Moreover, it helps distinguish closely packed and occluded NPs, providing valuable insights into their spatial arrangements and interactions. 

\subsection{Nanoparticle analysis}
We demonstrate the effectiveness of our AI-assisted particle analysis workflow on a diverse set of metallic NPs dispersed on catalytic support materials. Figure 2 shows object detection and segmentation results from BF-TEM images of PtCo NPs on carbon black, as well as HAADF STEM images of Cu NPs on silica and Ru NPs on $\gamma$-alumina, all captured from large field-of-view TEM images. The total processing time, including both detection and segmentation stages, is approximately one minute for 748 PtCo NPs and 339 Ru NPs, and 30 seconds for 173 Cu NPs. 

In BF-TEM, electron diffraction primarily drives contrast formation, whereas in HAADF STEM, contrast is influenced by variations in composition and sample thickness. The contrast modulation observed in the BF-TEM images of NPs arises from changes in crystal orientations relative to the electron beam. Image interpretation is further complicated by the contributions from the support material to background contrast and varying degrees of NP overlap. Using our proposed AI-assisted particle analysis workflow, 748 NPs with an average particle diameter of 4 $\pm$ 1.0 nm (mean $\pm$ standard deviation) were successfully detected and segmented from the BF-TEM image of PtCo NPs on carbon black support material, as shown in Figure 2 (top).

In HAADF STEM, image contrast is monotonic, with intensities increasing as both the sample thickness and atomic number (Z) of the elements increase. While a higher Z enhances the contrast, a significant increase in thickness can considerably reduce it. Consequently, identifying regions containing higher-Z elements against a background of relatively thin TEM foil and lighter support material is a more feasible task.   

STEM imaging presents significant challenges due to abrupt thickness variations in the underlying catalytic support material. The support materials often form micron-thick large aggregates decorated with planar surface facets and a dense network of pores and channels. Such geometric variations complicate contrast formation and make it difficult to trace the boundaries between the NPs and the surrounding background, particularly when there is a sudden increase in the support materials’s thickness. However, our proposed workflow successfully overcomes these challenges, as demonstrated by the detection and segmentation of hundreds of  Cu NPs (Figure 2, middle) and Ru NPs (Figure 2, bottom) with average particle diameters of 1.6 $\pm$ 0.2 nm and 2.9 $\pm$ 1.1 nm , respectively, from wide field-of-view HAADF STEM images.

The segmentation performance was evaluated using standard semantic segmentation metrics, as detailed in Table 1, to measure the overall overlap between the predicted and ground truth segmentations. The segmentation of the NPs, guided by bounding box prompting, achieved an average F1 overlap score of 0.91 ($\pm$ 0.01) across three catalytic systems: Cu NPs on silica, Pt NPs on $\gamma$-alumina, and PtCo NPs on carbon black datasets. Notably, the sparse distribution of the metallic NPs over the background and catalytic support material introduces an unbalanced representation of the data in the TEM images. This class imbalance, dominated by an overwhelming number of easy examples from the background, skews the loss function during the model training in CNNs, limiting their ability to focus on the minority foreground class. The high F1 score, precision, and recall values across each dataset demonstrate the effectiveness of our proposed two-stage workflow. First, it identifies the NPs in these imbalanced datasets using a multi-scale feature fusion and prediction approach and then segments them using a transformer model.
\begin{figure}[h]
	\centering
	\includegraphics[width=0.85\textwidth]{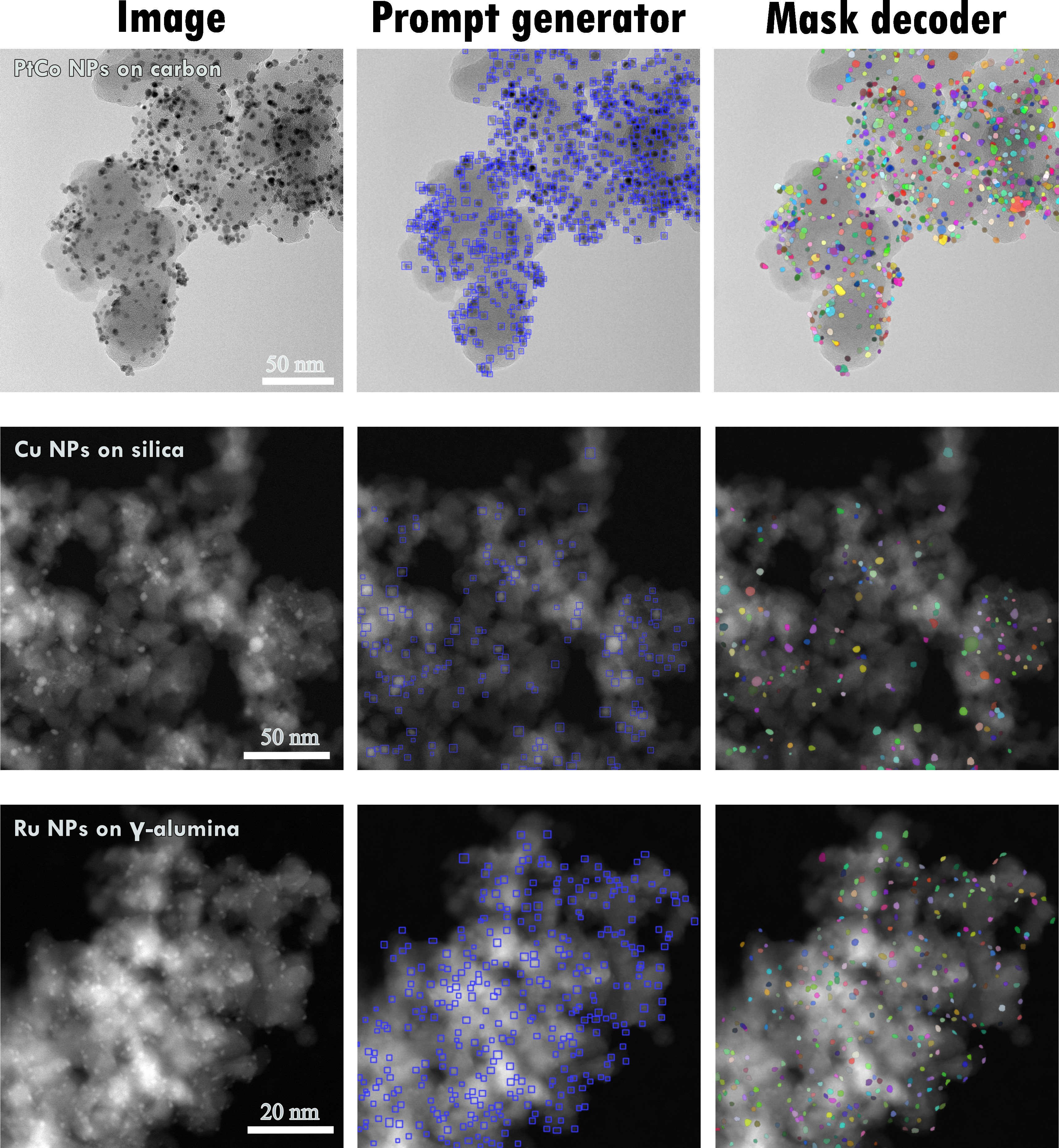}
	\caption{\small Sets of images showing object detection and segmentation results from BF-TEM images of PtCo NPs on carbon black (top), HAADF STEM images of Cu NPs on silica (middle), and Ru NPs on $\gamma$-alumina (bottom).}
	\label{fig:fig2}
\end{figure}
\setlength{\tabcolsep}{6pt} 
\renewcommand{\arraystretch}{1.2} 
\begin{table}
	\centering
	\caption{Segmentation-overlap results with corresponding standard deviations.}
	\begin{tabular}{ccccc}\\
		Material & Precision & Recall & F1 score & IoU \\
		\hline
		Pt NPs-$\gamma$-alumina &  0.89 $\pm$ 0.02 & 0.94 $\pm$ 0.01 & 0.92 $\pm$ 0.01 & 0.84 $\pm$ 0.02 \\
		Cu NPs-silica &  0.88 $\pm$ 0.03 & 0.91 $\pm$ 0.02 & 0.90 $\pm$ 0.01 & 0.81 $\pm$ 0.01 \\
		PtCo NPs-carbon&  0.93 $\pm$ 0.01 & 0.90 $\pm$ 0.01 & 0.91 $\pm$ 0.01 & 0.84 $\pm$ 0.01 \\
	\end{tabular}
	\label{tab:table1}

\end{table}
\setlength{\tabcolsep}{6pt} 
\renewcommand{\arraystretch}{1.2} 

\begin{table}
	\centering
	\caption{Boundary-overlap results with corresponding standard deviations.}
	\begin{tabular}{ccccc}\\
		Material & HD (nm) & RHD (nm) & Pixel size (nm) \\
		\hline
		Pt NPs-$\gamma$-alumina &  6.4 $\pm$ 1.5 & 1.4 $\pm$ 0.6 & 0.4 $\pm$ 0.2 \\
		Cu NPs-silica &  7.8 $\pm$ 3.7 & 0.16 $\pm$ 0.05 & 0.05 $\pm$ 0.01 \\
		PtCo NPs-carbon&  3.8 $\pm$ 2.3 & 0.4 $\pm$ 0.1 & 0.07 $\pm$ 0.03
	\end{tabular}
	\label{tab:table2}
\end{table}
The results in Table 1 indicate a relatively lower overlap ratio for the segmentation of Cu NPs compared to Pt NPs, which are associated with a higher average Z of elements in the HAADF STEM images. This lower performance for Cu NPs can be attributed to their low contrast against the silica substrate, especially when the smaller NPs are distributed on the thicker regions of the silica support material, as shown in Figure 2 (middle). The uncertainty in these diffuse boundaries can introduce ambiguity in identifying and manually annotating the ground truth segmentations, making precise annotation difficult even for human experts. The observed issues with Cu NPs highlight potential improvements in handling low-contrast images.

To further investigate our segmentation results, we assessed the degree of surface match for the NPs using the Hausdorff distance (HD) metric. While the segmentation-overlap results provide insight into overall performance across the entire NP, they are often more influenced by the high number of true positives. The HD metric specifically targets boundary-match errors between the ground truth and predicted segmentations of NPs, offering a more detailed evaluation of segmentation accuracy, particularly at the NP boundaries. 

Table 2 presents the largest segmentation error based on the average HD and the 90th percentile of robust HD (RHD) values. The RHD measurement is crucial for understanding the impact of the outliers and noise on model performance and for guiding the accuracy of boundary match. The average HD is 7.8 $\pm$ 3.7 nm, with the largest error observed for Cu NPs, aligning with the lower overall segmentation-overlap values reported for this material in Table 1. Interestingly, the 90th percentile RHD values follow a different trend, indicating significantly lower values of segmentation error: 1.4  $\pm$ 0.6 nm for Pt NPs, 0.16  $\pm$ 0.05 nm for Cu NPs, and 0.4  $\pm$ 0.1 nm for PtCo NPs.   

Additionally, we observe that RHD values scale with an average of 3.5 $\pm$ 0.45 pixels for the Pt and Cu NPs imaged both using the HAADF STEM technique. This systematic discrepancy across two catalytic systems is mostly expected considering the origins of the image formation in HAADF STEM, which is directly interpretable and near-linear compared to BF-TEM imaging. In contrast, there is a relatively more pronounced segmentation error per pixel for the BF-TEM images of PtCo NPs at the 90th percentile of the RHD. Contributing factors to the BF-TEM image formation, such as low Z-contrast, surface delocalization, and Fresnel fringes in regions with heavily overlapped particles, introduce significant ambiguity, making it difficult to segment these NPs with high pixel accuracy.

Metallic catalyst particles often exhibit bimodal or multimodal size distributions rather than being uniform. Analyzing a large sample size enhances the accuracy of the average particle size calculation by reducing the impact of the outliers and anomalous  measurements. As the number of sampled NPs increases, statistical errors decrease, therefore improving the reliability of the data and making the reported size more representative of the entire NP population. This approach also enhances the reproducibility and comparability of the results.
\begin{figure}[h]
	\centering
	\includegraphics[width=\textwidth]{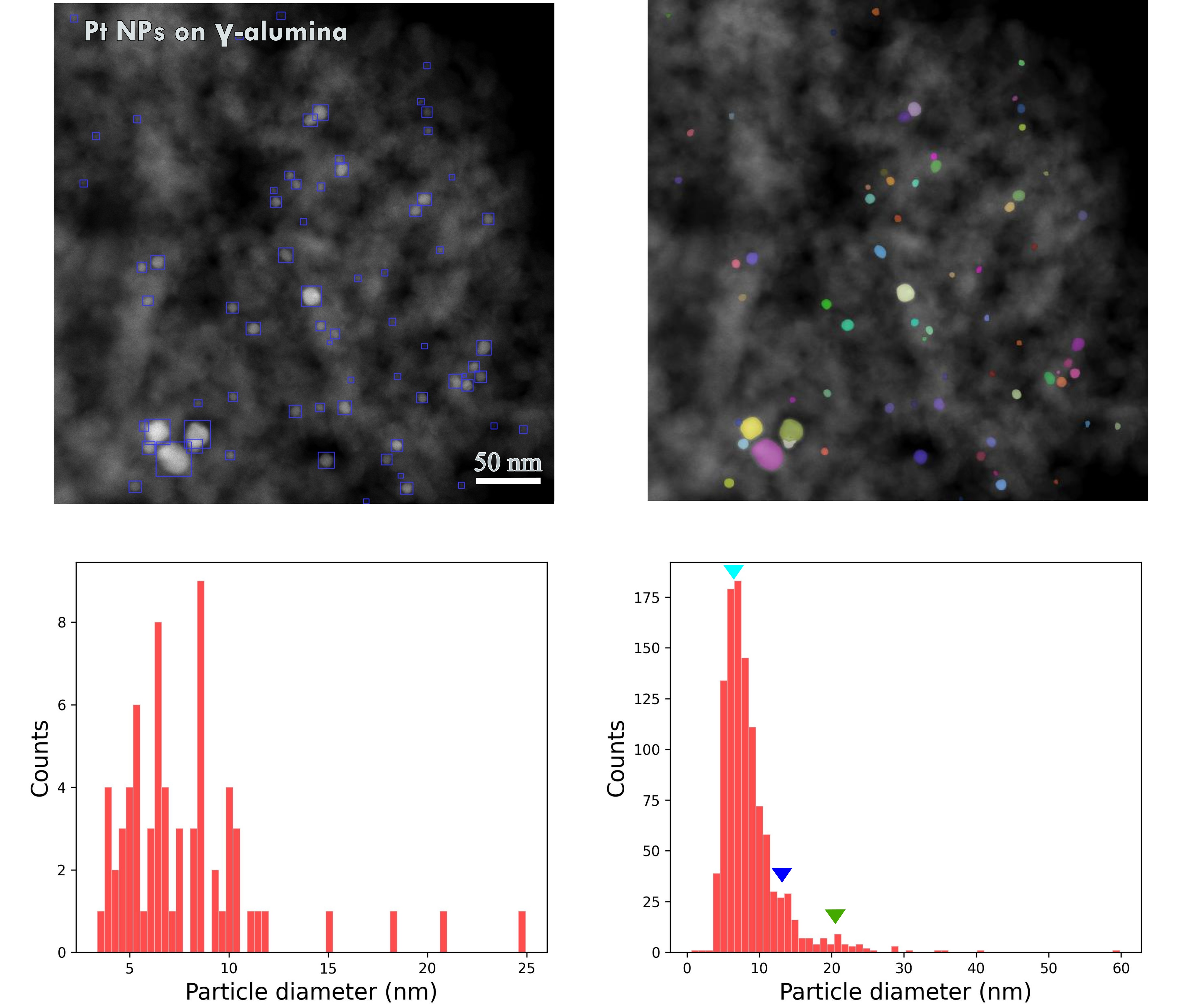}
	\caption{\small Object detection (top-left) and segmentation (top-right) results of Pt NPs on $\gamma$-alumina for 70 NPs. Side-by-side comparison of histograms generated using AI-assisted NP size analysis for the image containing 70 NPs (bottom-left) versus batch processing of 7 images containing 1,080 NPs (bottom-right). The total analysis times were 12 s and 212.7 s, respectively.}
	\label{fig:fig3}
\end{figure}

The particle size distribution histograms shown in Figure 3  illustrate the impact of sample size on the analysis of Pt NPs acquired using HAADF STEM. The histogram in Figure 3 (bottom-left) presents a particle size analysis from an image containing only 70 NPs with detection and segmentation results shown in Figure 3 (top). Figure 3 (bottom-right) represents an analysis of a larger sample of 1,080 NPs from a high-throughput analysis of 7 images. The batch image analysis of these 1,080 NPs was completed in 212.7 seconds, revealing more detailed insights into the particle size distribution. In addition to the prominent peak around 8 nm, a subtle shoulder around 15 nm and a broader peak around 20 nm in diameter are observed, as highlighted by the arrows. 

The smaller sample size results in a less distinct histogram in Figure 3 (bottom left). It poorly identifies the peak at 8 nm and fails to capture the additional size distribution characteristics present in the larger sample. This comprehensive analysis highlights the importance of larger sample sizes for accurately detecting and characterizing peaks in histograms, as they provide a more complete view of the NP size distribution and reduce the risk of misidentifying size characteristics.

When the NPs are connected or overlapped, particularly under reactive conditions, contrast similarities become more pronounced, complicating the segmentation of  NPs. Figure 4 illustrates how we tackled this issue using our two-stage approach of detection and segmentation for PtCo NPs with varying degrees of particle overlap. In this example, bounding box prompting guides SAM in localizing each NP instead of treating closely aligned NPs as a single entity. Leveraging the object detection model's ability to detect occluded objects, our AI-assisted NP analysis successfully disentangles four PtCo NPs into nine distinct NPs in a BF-TEM image. This small field of view resulted in more than a two-fold increase in the number of NPs identified on the support material. 

The detection model's anchor box-free, multi-scale prediction architecture enables the localization of NPs with varying degrees of overlap, morphologies, and sizes. Additionally, detection parameters, such as input image size, confidence level, and IoU threshold, can be interactively adjusted during inference to adapt to a new example dataset, providing greater flexibility and compatibility while managing the precision-recall trade-off. To further enhance its applicability, we plan to incorporate a broader range of catalytic particles with diverse shapes and contrast levels into the model. With recent advances in automatic data collection, AI-assisted NP analysis is positioned to revolutionize the field of TEM for routine quantitative measurement of catalysts on a statistically relevant scale, which is not currently possible.

\begin{figure}[t]
	\centering
	\includegraphics[width=1\textwidth]{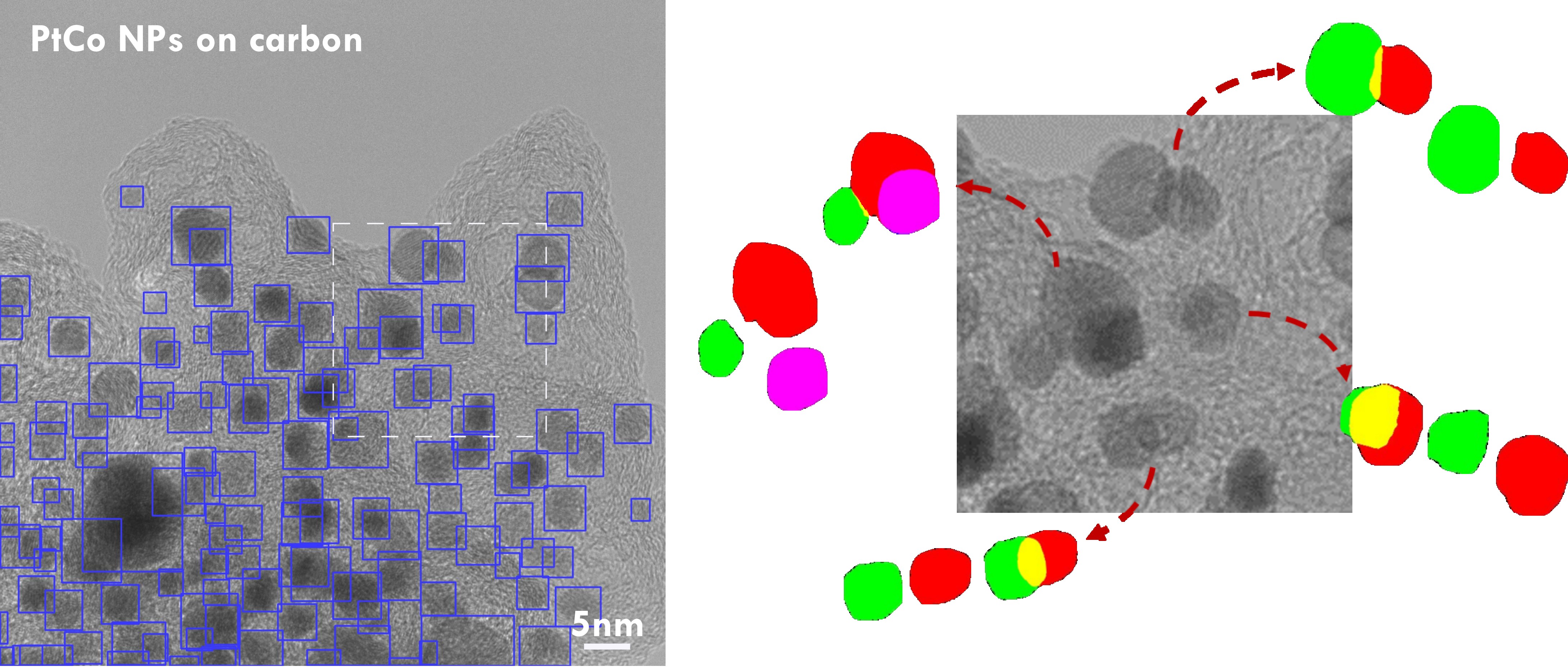}
	\caption{\small Object detection (left) and segmentation (right) of overlapped PtCo NPs. The AI-assisted particle analysis workflow disentangles four overlapping NPs into nine distinct NPs.}
	\label{fig:fig4}
\end{figure}
\section{Methods}
\subsection{Materials and data acquisition}
Our main goal is to assess the effectiveness of the prompt-guided zero-shot classification approach in the semantic segmentation of BF-TEM and HAADF STEM images of NPs while achieving a comprehensive high-throughput analysis. We acquired BF-TEM and HAADF STEM images of NPs on oxide and carbon catalytic support materials to accomplish this. The TEM samples were prepared by depositing a solution containing well-dispersed NPs on a lacey carbon film. 20$\%$ wt. PtCo (1:1 atomic ratio) on carbon black (Vulcan) powder was obtained from the FUELCELL store. A detailed description of the synthesis of Ru, Pt, and Cu NPs has been reported in an earlier paper  \cite{Hannagan2021}. We utilized a probe aberration-corrected 200 kV Thermo Fisher Scientific Spectra S/TEM microscope to acquire the BF-TEM and HAADF STEM images. HAADF STEM images were collected at the detector inner collection angle of 40 mrad, a beam current of 20 pA with a 0.1 nm probe size, and an accelerating voltage of 200 kV. To extend the depth of focus of the electron beam during STEM acquisitions, we adjusted the convergence angle of the illumination system to 15 mrad using the three-condenser lens optics of the microscope. 

The Thermo Fisher Scientific Velox software platform was utilized to acquire BF-TEM and HAADF-STEM images. We employed a custom Python script to read the Velox files containing image data and metadata, integrating this metadata for pixel size calibration and microscope settings. Images for AI-assisted NP size analysis were loaded without any pre-processing steps, such as de-noising, smoothing, or edge enhancement operations. The segmentations were converted into a batch of binary images and analyzed using the Scikit-image Python library to extract relevant properties.

\subsection{Training and optimization}
For object detection model training, we used mini-batch gradient descent with a batch size of 32 and a stochastic gradient descent (SGD) optimizer with weight decay. Default hyperparameters from the Ultralytics deep learning framework were applied for gradient averaging and updating first and second moments. The YOLOv8x model weights were initialized using transfer learning from weights pre-trained on the large-scale MS COCO (Microsoft Common Objects in Context) dataset comprising 330,000 annotated images and 80 classes. For fine-tuning the YOLOv8x model, we selected a total of 25 ground-truth images from sets of Pt NPs on $\gamma$-alumina, Cu NPs on silica, Pd NPs on $\gamma$-alumina, and PtCo NPs on carbon black of HAADF STEM and BF-TEM images. The corresponding ground truth bounding boxes were manually annotated and divided into training and validation datasets. 

We applied rotation, vertical and horizontal flipping, brightness adjustments, and noise transformations to generate a diverse set of images capturing variations in the location and shape of the features. With data augmentation, a total of 75 images, and their annotations were used for model training and 3 images for validation. The complete training of the YOLOv8x model was performed at an input resolution of 640x640 pixels using NVIDIA A100 GPUs located at the California Nanosystems Institute (CNSI), University of California Santa Barbara.

\subsection{Evaluation metrics}
Evaluation metrics are crucial for assessing the performance of the network and establishing the model for automated semantic segmentation. We evaluated the fidelity of our segmentation work using standard semantic segmentation metrics: Intersection over Union (IoU), F1 score, recall (the proportion of actual positives correctly identified), and precision (the proportion of true positives among all positive predictions). These metrics, precision, recall, IoU, and F1 score, are derived from the confusion matrix and defined as follows:\\
\begin{equation}
	Precision =\frac{TP}{TP+FP}
\end{equation}
\begin{equation}
	Recall=\frac{TP}{TP+FN}
\end{equation}\\
\begin{equation}
	IoU=\frac{TP}{TP+FP+FN}
\end{equation}\\
\begin{equation}
	F1 score =\frac{2xPrecisionxRecall}{Precision+Recall}
\end{equation}\\

where true positives (TP), false positives (FP), and false negatives (FN) represent per pixel classifications of the confusion matrix.

The F1 score is calculated as the harmonic mean of precision and recall. It combines both metrics into a single score, ensuring that precision and recall are equally considered even when the data is imbalanced. 

We also evaluated our semantic segmentation results by measuring the dissimilarities at the segmentation boundaries. Hausdorff distance (HD) is a boundary distance-based metric that measures the largest segmentation error in the overlap between the ground truth and predicted segmentations \cite{huttenlocherComparingImagesUsing1993}. Given two sets of points A and B, HD distance is defined as:\\
\begin{equation}
	H D(A, B)=\max (h d(A, B), h d(B, A))
\end{equation}\\
where hd(A,B) and hd(B,A)  are directed Hausdorff distances:\\
\begin{equation}
	h d(A, B)=\max _{a \in A} \min _{b \in B}\|a-b\|
\end{equation}\\
\begin{equation}
	h d(B, A)=\max _{b \in B} \min _{a \in A}\|a-b\|
\end{equation}\\
The functions hd(A,B) and hd(B,A)  measure the distances between points in sets A and B, that are farthest from their nearest neighbors, with HD (A,B) giving the largest of these distances. The notation ‖a-b‖ denotes the Euclidean norm between points a in A and b in B. A well-documented characteristic of the HD is its sensitivity to outliers and noise  \cite{maiseliHausdorffDistanceOutliers2021,huttenlocherComparingImagesUsing1993,noauthor_surface_2022}; thus, we report robust HD (RHD) values considering the percentile of the largest segmentation errors along with the maximum HD. By measuring the RHD values, we aim to mitigate the impact of outliers and noise on the HD metric.

We used the mean average precision (mAP) metric to evaluate the bounding box predictions. The calculation of mAP involves predicting bounding boxes on the validation dataset and is defined as follows ;\\
\begin{equation}
	mAP=\ \frac{1}{n}\sum_{C=1}^{C=n}{AP}_C 
\end{equation}

Steps for calculating mAP:
\begin{itemize}
	\item \emph{Generate model predictions:} Apply various confidence thresholds to obtain model predictions. 
	\item \emph{Filter predictions:} Utilize an IoU threshold to filter the predictions based on the bounding box overlap ratio.
	\item \emph{Construct Precision-Recall Curve:} For each class, create a precision-recall curve based on the filtered predictions.
	\item \emph{Compute AP:} Calculate the AP by integrating the area under the precision-recall curve.
	\item \emph{Compute the mAP:} Calculate the mAP values across all classes; in our study, we focus on one class.
\end{itemize}
mAP is calculated on the validation dataset at the end of each epoch during training. The model checkpoint corresponding to the epoch that achieves the highest mAP score is selected for object detection inference. 

We report a mAP@0.5 (IoU at 0.5) of 0.78 and a mAP@0.5:0.95 (IoU from 0.5 to 0.95 in increments of 0.05) of 0.41 on the validation data for the object detection model. Overlapping particles can complicate the mAP calculation with stricter IoU constraints, which accounts for the discrepancy between the mAP@0.5 and mAP@0.5:0.95 scores.  

For the evaluation of semantic segmentation, we manually annotated 226 NPs for PtCo on carbon black, 136 NPs for Cu on silica, and 140 NPs for Pt on the $\gamma$-alumina datasets. We measured precision, recall, F1, and IoU scores for each dataset, comparing these overlap metrics between the hand-annotated segmentations and the bounding box prompt-guided segmentations generated by SAM.

\section{Conclusion}
We present a two-stage AI-assisted particle analysis workflow that integrates a prompt generator model for identifying regions containing NPs and zero-shot learning for precise segmentation. This automated workflow enables scientists and engineers to apply it directly to their NP datasets without the need for model retraining. Our results demonstrate that the object detection model effectively detects small NPs dispersed on the catalytic support materials within a wide field-of-view of the TEM images and performs well even when NPs vary significantly in size or are overlapped. This integrated methodology improves the efficiency, reliability, and scalability of NP analysis. 

\section*{Data Availability}
Python code is publicly available at the link below.
\section*{Code Availability}
The code for this study, including the model weights and a demonstration script, is publicly available online on GitHub $\href{https://github.com/ArdaGen/STEM-Automated-Nanoparticle-Analysis-YOLOv8-SAM}{S/TEM Automated Nanoparticle Analysis.}$

\section*{Competing interests}
The authors declare no competing interests.
\section*{Correspondence}
Correspondence and requests for materials should be addressed to $\href{mailto:ardagenc@ucsb.edu}{ardagenc@ucsb.edu}$
\section*{Acknowledgement}
Computational facilities (CNS-1725797) provided by the Center for Scientific Computing (CSC), which is operated by the California NanoSystems Institute and the Materials Research Lab (MRSEC; NSF DMR 2308708) at UCSB. LK was supported by the U.S. Department of Energy, Office of Science, Basic Energy Sciences, Chemical Sciences, Geosciences, and Biosciences Division, Catalysis Science program, FWP 47319. We thank Profs. Raphaele Clement and Ram Seshadri for providing the PtCo on Vulcan powder material.

\bibliographystyle{ieeetr}
\bibliography{zotero}

\end{document}